\documentclass{article}
\usepackage{color}
\usepackage{graphicx}
\usepackage{amsmath, amsthm}
\usepackage{mathtools}
\usepackage{natbib}
\usepackage{url}
\RequirePackage[colorlinks,citecolor=blue,urlcolor=blue]{hyperref}
\usepackage{xcolor}
\usepackage{chngcntr}
\usepackage{subfig}
\usepackage{hyperref}
\usepackage{algorithm}
\usepackage[noend]{algpseudocode}
\usepackage{setspace}

\usepackage{xfrac}
\usepackage{multirow}
\usepackage{amsfonts}

\usepackage{enumitem}
\usepackage{caption}
\usepackage{array}
\usepackage{makecell}

\evensidemargin=\oddsidemargin
\usepackage{etoolbox}

\newif\ifabbreviation
\pretocmd{\thebibliography}{\abbreviationfalse}{}{}
\AtBeginDocument{\abbreviationtrue}

\begin{document}
	\newcommand{\bb}{\boldsymbol{\beta}}

	\title{COBRA-DOSE: Copula-based Bayesian Model Averaging for Dose Selection}

	%% Group authors per affiliation:

	\author{Luke Hagar\footnote{Luke Hagar is the corresponding author and may be contacted at \url{l.hagar@uq.edu.au}.} \hspace{35pt} Min Zhang$^*$ \hspace{35pt} Ranjeny Thomas$^{\dagger}$ \hspace{35pt} Andrew J. Martin$^*$ \bigskip \\ 
 $^*$\textit{Clinical Trials Capability, The University of Queensland} \\ $^{\dagger}$\textit{Frazer Institute, The University of Queensland}}

	\date{}

	\maketitle

	\begin{abstract}

Early-phase clinical trials for dose selection typically enrol few patients and aim to identify doses that are both safe and promising for further study. While traditional approaches identify the maximum tolerated dose, modern trials for targeted therapies often seek the optimal biological dose, defined as the lowest dose achieving sufficient biological activity with acceptable safety. In immunology settings, assessment of biological activity is based on multiple biomarkers or clinical endpoints. Clinicians leading dose-selection efforts would thus benefit from transparent summaries of the probabilities of observing combinations of biomarker outcomes across doses. However, such inference is challenging in small samples where complex modelling assumptions are difficult to verify. To address this limitation, we propose COBRA-DOSE, a framework for posterior predictive inference based on two endpoints that models dependence via copulas and accounts for uncertainty in both marginal distributions and dependence structures through Bayesian model averaging. This approach avoids reliance on a single model and yields interpretable quantities for clinical decision making. We demonstrate the performance of COBRA-DOSE using DEN-181, a phase I immunology trial in rheumatoid arthritis. We also provide a general implementation of our approach through the \texttt{CobraDose} package in R.

		\bigskip

	\end{abstract}

	\maketitle

	%\pagebreak
	\baselineskip=19.5pt

	%%%%%%%%%%%%%%%%%%%%%%%%%%%%%%%%%%%%%%%%%%%%%%%%%%%%%%%%%%%%%%%%%%%%%%%%%%%%%%%%%%%%%%%%%%%%%%%%%%%%%
	%%%%%%%%%%%%%%%%%%%%%%%%%%%%%%%%%%%%%%%%%%%%%%%%%%%%%%%%%%%%%%%%%%%%%%%%%%%%%%%%%%

	%%%%%%%%%%%%%%%%%%%%%%%%%%%%%%%%%%%%%%%%%%%%%%%%%%%%%%%%%%%%%%%%%%%%%%%%%%%%%%%%%%%%%%%%%%%%%%%%%%%%%
	%%%%%%%%%%%%%%%%%%%%%%%%%%%%%%%%%%%%%%%%%%%%%%%%%%%%%%%%%%%%%%%%%%%%%%%%%%%%%%%%%%%%%%%%%%%%%%%%%%%%%

    \section{Introduction}\label{sec:intro}

    Robust dose-finding methodologies are critical in early-phase clinical trials to support safe and efficient dose selection for further development. Regulatory guidance has long established that dose selection should be informed by the relationship between dose and both beneficial and undesirable effects \citep{FDA1994ICHe4, FDA2003guidance}, with the criteria for how this is best achieved depending on the therapeutic context. Designs for cytotoxic agents in oncology historically aimed to identify the maximum tolerated dose, based on the assumption that efficacy increases with the dose until toxicity becomes unacceptable \citep{hryniuk1988more, storer1989design}. More recently, attention has focused on identifying the optimal biological dose (OBD), typically defined as the lowest dose achieving sufficient biological activity with acceptable safety \citep{thall1998strategy,le2009dose, fraisse2021optimal}. This focus has been reinforced by regulatory guidance that advocates for the incorporation of a wide range of data beyond toxicity into dose recommendations \citep{fda2024optimizing}.

    The OBD-targeting principle extends to other settings, including early-phase immunology trials, which are the focus of this paper. Particularly in tolerance-inducing immunotherapies, severe acute toxicity is less likely to be the primary constraint on dose selection. Moreover, the dose-response relationship for pharmacodynamic endpoints may be non-monotone: excessively high doses can suppress rather than promote the desired immune response. Pharmacodynamic evidence therefore plays a central role in identifying doses for further evaluation \citep{todd2016regulatory, nikolic2022tolerogenic} using frameworks such as MCP-Mod \citep{bretz2005combining}. Where a single biomarker is insufficient to  fully capture the pharmacodynamic response, evidence from multiple endpoints must be jointly considered. Jointly incorporating information from these endpoints into statistical inference for dose selection is, however, challenging given the small sample sizes inherent to early-phase trials.
   
    One strategy to address this challenge involves specifying a joint model across the endpoints (see e.g., \citet{thall2004dose,yin2006bayesian}). While such approaches are useful, they may involve restrictive assumptions about the dependence structure or dichotomise information from continuous outcomes, which can lead to a loss of information. An alternative strategy is to synthesise information from multiple endpoints into a single measure using a utility function \citep{tao2015dose, d2025utility}. This approach can be used to circumvent strict assumptions on the dependence structure between endpoints. However, defining an appropriate utility function may present challenges for clinicians, and numeric summaries of the utility function may be difficult to interpret. An approach that accommodates uncertainty in the joint distribution without sacrificing the interpretability of its outputs would therefore address the limitations of both strategies.
    
    We develop such an approach for dose-selection trials with two endpoints in this paper. Our proposed method leverages copulas \citep{nelsen2006introduction} and Bayesian model averaging (BMA) \citep{hoeting1999bayesian, fragoso2018bayesian} to accommodate uncertainty in the joint distribution used for data analysis. We refer to this approach as COBRA-DOSE: Copula-based Bayesian model aveRAging for DOSE selection. The posterior predictive distribution \citep{rubin1984bayesianly,gelman1996posterior}, which characterises the distribution of data for future patients while naturally propagating uncertainty from the current posterior, gives rise to interpretable data summaries that are particularly well suited to the small-sample setting of early-phase trials. These summaries include the posterior predictive probabilities associated with joint events that correspond to various combinations of the two outcomes. Collections of these posterior predictive probabilities can be compared across doses to guide dose recommendations.  Our proposed approach can be broadly implemented using the \texttt{CobraDose} package in R \citep{hagar2026cobra}. 

    Our proposed methodology is illustrated in the context of DEN-181 \citep{sonigra2022randomized}, a phase I dose-finding trial that explores the impact of a single dose of a new immune-modifying treatment for rheumatoid arthritis. Various metabolic and immunologic biomarkers are relevant for dose recommendations in this context. In early-phase drug development, single-dose dose-ranging studies are commonly followed by multiple-dose studies. Posterior predictive inference from the single-dose trial can therefore help inform dose selection for the future follow-up multiple-dose studies. While our work here is illustrated utilising DEN-181, it can be applied within a broader class of early-phase immunology trials involving multiple biomarkers. 

     The remainder of this article is structured as follows. Section \ref{sec:den} introduces the DEN-181 trial that serves as this paper's illustrative example. In Section \ref{sec:methods}, we propose our approach for posterior predictive inference in dose-finding trials based on bivariate modelling with copulas, hierarchical modelling across doses, and BMA. We conduct numerical studies in Section \ref{sec:sim} to calibrate our prior distributions for small-sample analysis. We apply our proposed approach to the data from DEN-181 in Section \ref{sec:real}. In this context, we demonstrate the value our methodology provides over simpler, univariate approaches that model each endpoint separately. Section \ref{sec:code} overviews the R package created to implement our proposed methods. In Section \ref{sec:disc}, we conclude with a summary and discussion of extensions to this work.

    \section{Illustrative Example}\label{sec:den}

    DEN-181 \citep{sonigra2022randomized} was a phase I randomised, double-blind, placebo-controlled trial in adult patients with anti-citrullinated protein antibody+ (ACPA+) rheumatoid arthritis (RA) on stable treatment with methotrexate. The investigational product is an immunomodulatory therapy based on liposomes containing collagen II peptide and calcitriol, intended to promote antigen-specific immune tolerance in ACPA+ RA. The study was a single-dose escalation trial across $J=3$ active dose levels (indexed by $j = 1, 2,3$) plus placebo $(j = 0)$. Data from 16 patients are available: $n_0= 5$ patients in the placebo group, and $n_1 = 4$, $n_2 = 3$, and $n_3 = 4$ in the active dose groups. While the DEN-181 phase I trial assessed the effect of a single dose, further evaluation of repeat dosing in RA patients was identified as an important future direction \citep{sonigra2022randomized}. Our methodology enables this transition: the posterior predictive distribution based on the single-dose pharmacodynamic data from DEN-181 can be used to help inform dose selection for a future repeat-dosing trial in a related RA population.

     We later apply COBRA-DOSE to DEN-181 data as a retrospective statistical analysis of dose recommendation. Of the broader set of biomarker measurements collected in DEN-181, our present analysis focuses on the two continuous biomarkers for illustration. 
     
    The first biomarker, total T cell count ($k = 1$), provides a broad readout of T cell immunity, and the second biomarker, inflammatory dendritic cell count ($k = 2$), is a more specific readout of an inflammatory antigen-presenting cell population. Both measurements are positive and can take non-integer values. As detailed in Section \ref{sec:real}, these biomarkers are summarised relative to baseline for analysis. Because they capture related but distinct aspects of pharmacodynamic response and may be statistically dependent, it would be ideal if dose recommendation were based on their joint behaviour rather than on either biomarker in isolation. The proposed method provides a structured quantitative basis for incorporating both biomarkers into dose recommendation, which is particularly valuable in small-sample settings such as DEN-181.

    \section{Methodology}\label{sec:methods}

    \subsection{Bivariate Modelling with Copulas}\label{sec:methods.cop}

    In this paper, we focus broadly on dose-finding trials with two endpoints indexed by $k = 1, 2$. The trial explores $J+1$ doses with labels $j = 0, 1, \dots, J$, such that dose $j = 0$ corresponds to a placebo. We suppose that $y_{kji}$ represents a datum corresponding to endpoint $k$ for patient $i = 1, \dots, n_j$ randomised to dose $j$. The relative changes in multiple biomarkers over the course of the trial are often of interest in immunology settings. For such trials, $y_{kji} = y_{kji}^*/y_{kji}^b$, where $y_{kji}^b$ and $y_{kji}^*$ are respectively the values of endpoint $k$ at baseline and at the end of follow-up for patient $i$ receiving dose $j$.

    We first introduce our bivariate modelling approach in the context of a single dose $j$. We thus retain only the subscript for the endpoint $k$ in the remainder of this subsection. Our modelling approach considers the joint distribution of the random variables $\boldsymbol{Y} = (Y_1, Y_2) \in \mathbb{R}^2$. This joint behaviour is characterised by the joint distribution function $H(\boldsymbol{y})$. Each endpoint in $\boldsymbol{Y}$ also has a marginal cumulative distribution function (CDF) denoted by $F_k(y_k) = Pr(Y_k \le y_k), ~ k = 1, 2$. The endpoints explored in our illustrative example correspond to biomarkers that are continuous and positive. For such trials, suitable choices for the marginal distributions of $Y_k \in \mathbb{R}_{>0}$ may include gamma or lognormal distributions.  

  Copulas flexibly allow for the dependence structure of $\boldsymbol{Y}$ to be considered separately from its marginals. Copulas can be applied in settings with an arbitrary number of endpoints, but we introduce copulas within the bivariate context explored in this paper. We let $U_1$ and $U_2$ be uniformly-distributed random variables over the unit interval $[0,1]$. The distribution function
	\begin{equation*}\label{eqn:copula}
		C(u_1, u_2) = Pr(U_1 \le u_1, U_2 \le u_2)
	\end{equation*} 
is such that $C: [0,1]^2 \rightarrow [0,1]$ is a copula \citep{nelsen2006introduction}. Sklar's theorem \citep{sklar1959fonctions} clarifies the relationship between the copula $C$, the multivariate joint CDF $H(\boldsymbol{y})$, and the univariate marginal CDFs $F_k(y_k)$ for $j = 1, 2$:
	\begin{equation*}\label{eqn:sklar}
		H(\boldsymbol{y}) = C(F_1(y_1), F_2(y_2)).
	\end{equation*} 

   A copula is absolutely continuous \citep{nelsen2006introduction} if and only if it has a density function $c(u_1, u_2)$ such that
	\begin{equation*}\label{eqn:copula_den}
		C(u_1, u_2) = \int_{0}^{u_1} \int_{0}^{u_2} c(v_1,  v_2) dv_1 dv_2.
	\end{equation*} 
In this paper, we require that $C$ is absolutely continuous to write the joint probability density function (PDF) of $\boldsymbol{Y}$ as
\begin{equation*}\label{eqn:copula_pdf}
		h(\boldsymbol{y}) = f_1(y_1) \times f_2(y_2) \times c(F_1(y_1), F_2(y_2)),
	\end{equation*} 
where $f_k(y_k)$ is the continuous marginal PDF of $Y_k$. So as to not introduce arbitrary restrictions on the support of $\boldsymbol{Y}$, we also require the copula $C$ to have full support over $[0,1]^2$.  

We consider four possible families for the copula in this paper: independence, Clayton \citep{clayton1978model}, Gaussian \citep{clemen1999correlations}, and Gumbel \citep{gumbel1960bivariate}. The variables $Y_1$ and $Y_2$ are independent under the independence copula. Apart from the independence copula, each family is defined via a parameter $\theta \in \mathbb{R}$ that characterises the strength and direction of dependence. The strength and direction of dependence can also be summarised via Kendall's $\tau \in [-1,1]$ \citep{kendall1938new}, which measures rank correlation in terms of how similar the orderings of bivariate data are when ranked by each endpoint. Kendall's $\tau$ values of 1 and -1 respectively define settings where two random variables exhibit perfect positive and negative dependence. When two variables have no monotonic dependence, Kendall's $\tau$ is 0.  

In theory, the Clayton and Gaussian copulas accommodate both positive and negative dependence between $Y_1$ and $Y_2$, whereas the Gumbel copula only accommodates positive dependence. The four copulas, along with their admissible $\theta$ values and Kendall's $\tau$ values as a function of $\theta$, are summarised in Table \ref{tab:cop}. To define the bivariate Gaussian copula, we let $\Phi(y)$ be the CDF of the standard normal distribution. We also let $\Phi_{\theta}(y_1, y_2)$ be the joint CDF of the bivariate normal distribution with mean $\boldsymbol{0}$, unit variances, and correlation $\theta$. We note that the Clayton and Gaussian copulas approximate the independence copula as $\theta \rightarrow 0$; the Gumbel copula approaches the independence copula as $\theta \rightarrow 1^+$.

\begin{table}[!tb]
\centering
\caption{Summary of the admissible parameters and Kendall's $\tau$ for the four copula families considered}
\label{tab:cop}
\begin{tabular}{cccc}
Family       & $C_{\theta}(u_1, u_2)$        & Parameter & Kendall's $\tau$ \\ \hline
Independence & $u_1u_2$   & ---       & 0                \\
Clayton      & $\left[\max\{u_1^{-\theta} + u_2^{-\theta}, 0\}\right]^{-1/\theta}$ & $\theta \in [-1, \infty)\setminus \{0\}$         & $\theta/(\theta+2)$               \\[2pt]
Gaussian     &  $\Phi_{\theta}(\Phi^{-1}(u_1), \Phi^{-1}(u_2))$          &      $\theta \in [-1,1]$     & $2\sin^{-1}(\theta)/\pi$                 \\[2pt]
Gumbel       &  $\exp\left[- ((-\log(u_1))^{\theta} + (-\log(u_2))^{\theta})^{1/\theta}\right]$          & $\theta \in [1, \infty)$          & $1 - 1/\theta$                
\end{tabular}
\end{table}

\begin{figure}[!tb] \centering 
		\includegraphics[width = 0.8\textwidth]{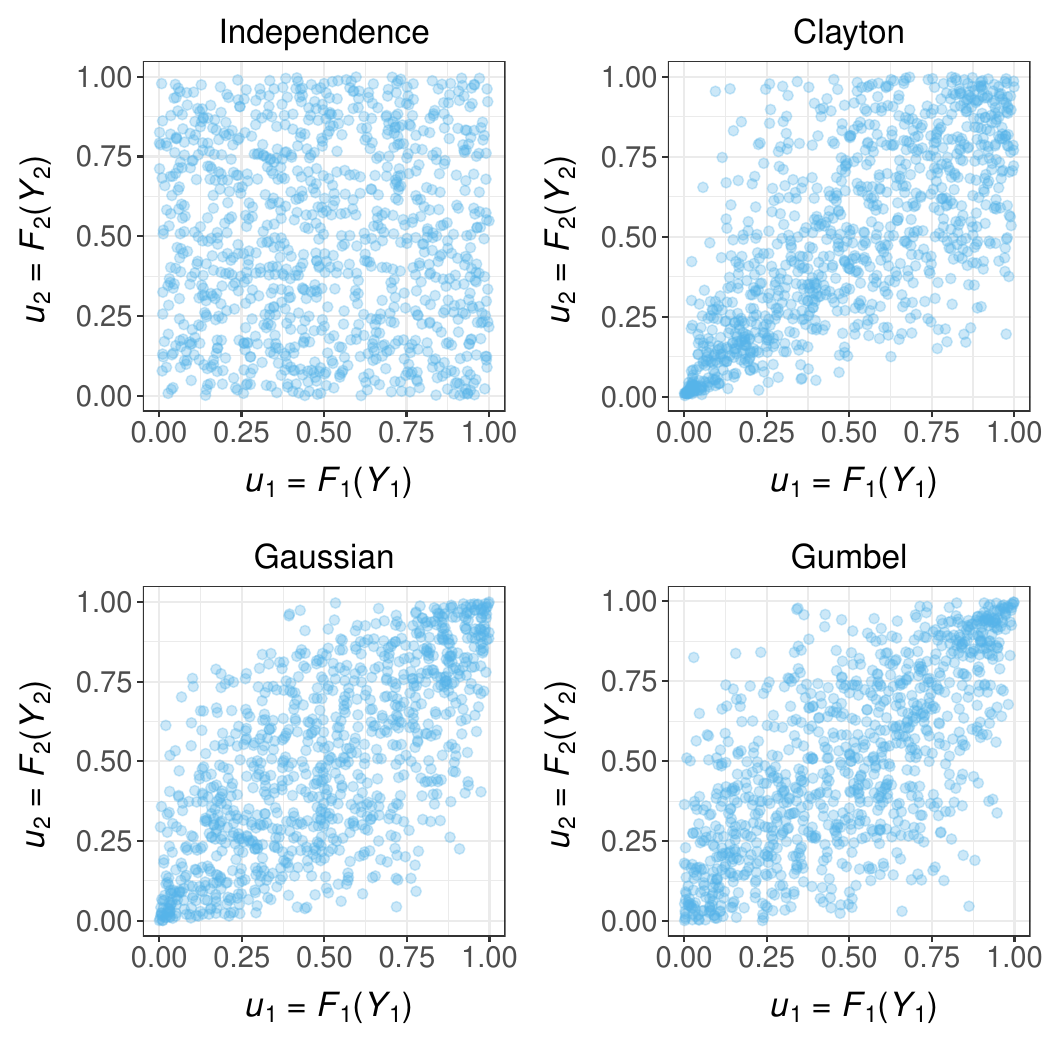} 
		\caption{\label{fig:cop} Samples of 1000 points from four copula families, where Kendall's $\tau$ is 0.5 for all non-independence copulas.} 
	\end{figure}

    We further illustrate the differences between the four copula families in Figure \ref{fig:cop}. This figure compares random samples of 1000 points in $[0,1]^2$ from each family. To ensure the same magnitude and direction of dependence across the non-independence copulas, we consider Clayton, Gaussian, and Gumbel copulas where $\theta$ is selected so that Kendall's $\tau$ in Table \ref{tab:cop} is 0.5. As expected, the points from the independence copula appear to be randomly scattered over $[0,1]^2$. The samples from the other families appear to cluster around the line $u_1 = u_2$, with tail behaviour differing between families. The Clayton copula accommodates lower tail dependence since the points are tightly clustered in the bottom left corner of its subplot. The Gaussian copula induces symmetric dependence in the lower and upper tails, with no tail dependence in either. The Gumbel copula favours upper tail dependence as its points are more concentrated in the top right corner. Ultimately, considering these four copula families provides a tractable framework for capturing a range of dependence structures between the two endpoints.

    \subsection{Hierarchical Modelling}\label{sec:methods.hier}

    To maximise use of the limited data available in small dose-finding trials, we leverage a three-stage hierarchical model that borrows information across dose levels. We henceforth reintroduce the subscripts for the dose $j$ and patient $i$ when referring to the data $Y_{kji} \in \mathbb{R}_{>0}$. The bivariate data  $(Y_{1ji}, Y_{2ji})$ for patient $i$ receiving dose $j$ are jointly modelled using an approach that accounts for uncertainty in the choice of marginal distributions and copula. We define categorical variables $\{Z_k\}_{k=1}^2 \in \{1,2\}$ and $Z_c \in \{1, 2, 3, 4\}$ that correspond to different choices for the marginal distribution of the relative change in endpoint $k$ and the copula.
    
    We suppose the marginal distribution for endpoint $k$ is either gamma ($Z_k = 1$) or lognormal ($Z_k = 2$). While other distributions could be considered, these distributions are well suited to model data associated with relative changes in continuous and positive biomarkers. We further suppose that the copula family is one of those introduced in Section \ref{sec:methods.cop}: independence ($Z_c = 1$), Clayton ($Z_c = 2$), Gaussian ($Z_c = 3$), or Gumbel ($Z_c = 4$). Across all combinations of marginal distributions and copulas, we obtain a set of 16 potential joint models. Each joint model corresponds to a distinct combination of $Z_1$, $Z_2$, and $Z_c$. We therefore allow the family for the marginal distribution to differ across endpoints, but we assume that the marginal and copula families are the same for all doses. 

    Nevertheless, we allow the \emph{parameters} of the gamma or lognormal distributions and copula to differ between doses according to a hierarchical model. To give rise to a common parametrisation across the gamma and lognormal distributions, we explicitly define the marginal distribution for each dose $j$ and endpoint $k$ in terms of its mean $m_{kj} \in \mathbb{R}_{> 0}$ and coefficient of variation $cv_{kj} \in \mathbb{R}_{> 0}$. We can easily convert between the mean and coefficient of variation and the standard parameters of the gamma and lognormal distributions. For the gamma distribution parametrised by shape $\alpha_{kj}$ and rate $\beta_{kj}$, we have that 
\begin{equation}\label{eqn:gamma}
		\alpha_{kj} = \dfrac{1}{cv_{kj}^2}~~~\text{and}~~~\beta_{kj} = \dfrac{1}{m_{kj}cv^2_{kj}}.
	\end{equation} 
    For the lognormal distribution parametrised by mean $\mu_{kj}$ and standard deviation $\sigma_{kj}$, it follows that
    \begin{equation}\label{eqn:lnorm}
		\sigma_{kj} = \sqrt{\log(1 + cv_{kj}^2)}~~~\text{and}~~~\mu_{kj} = \log(m_{kj}) - \frac{1}{2}\sigma_{kj}^2.
	\end{equation} 

    We note that the interpretation and support of the copula parameter $\theta$ changes according to its family. To similarly facilitate a common parametrisation across copula families, we define the copula in terms of Kendall's $\tau$. In this common parametrisation, we force Kendall's $\tau$ to be within $(0,1)$ for the non-independence copulas. Our approach thus only accommodates positive dependence between the two endpoints. We can, however, instead apply our proposed methods to $1/Y_{1}$ and $Y_{2}$ if $Y_1$ and $Y_2$ exhibit negative dependence on the untransformed scale. We discuss the implications of this transformation on posterior predictive inference in Section \ref{sec:methods.ppp}. Because we include the independence copula in our joint modelling approach, our methodology could still reasonably be applied when $Y_1$ and $Y_2$ exhibit weak negative dependence under some doses and weak positive dependence under others. Our approach does not accommodate settings with strong positive dependence in some doses and strong negative dependence in others, but hierarchical modelling would be problematic in this setting regardless. For the non-independence copulas, we can convert between Kendall's $\tau$ and the parameter $\theta$ by inverting the expressions in the final column of Table \ref{tab:cop}. Kendall's $\tau$ is not incorporated into models defined using the independence copula.

    Our three-stage hierarchical model that borrows information across dose levels $j = 0, 1, \dots J$ for endpoints $k =1, 2$ is defined below:
    \begin{equation}\label{eqn:hier}
\begin{aligned}
\text{Stage 1:}\quad 
&(Y_{1ji}, Y_{2ji}) \sim \mathrm{BIVAR}(z_1, m_{1j}, cv_{1j}, \; z_2, m_{2j}, cv_{2j}, \; z_c, \tau_j) \\[6pt]
\text{Stage 2:}\quad 
&m_{kj} \sim \mathrm{TN}(\mu_{mk}, \sigma_{mk}^2, 0, \infty), \quad 
cv_{kj} \sim \mathrm{TN}(\mu_{cvk}, \sigma_{cvk}^2, 0, \infty), \\
&\mathrm{logit}(\tau_j) \sim \mathcal{N}(\mu_\tau, \sigma_\tau^2), \\
&z_k \sim \mathrm{Multinomial}(1, \pi_k), \quad 
z_c \sim \mathrm{Multinomial}(1, \pi_c) \\[6pt]
\text{Stage 3:}\quad 
&\mu_{mk} \sim \mathcal{N}(1, 0.2), \quad 
\sigma_{mk} \sim \mathrm{TN}(0.2, 0.2, 0, \infty), \\
&\mu_{cvk} \sim \mathcal{N}(0.5, 0.2), \quad 
\sigma_{cvk} \sim \mathrm{TN}(0.1, 0.2, 0, \infty), \\
&\mu_\tau \sim \mathcal{N}(0.25, 0.5), \quad 
\sigma_\tau \sim \mathrm{TN}(0.25, 0.5, 0, \infty), \\
&\pi_k \sim \mathrm{Dir}(1.625, 1), \quad 
\pi_c \sim \mathrm{Dir}(1, 1, 1.05, 1),
\end{aligned}
\end{equation}
where $\mathrm{TN}(\mu, \sigma^2, 0, \infty)$ denotes the $\mathcal{N}(\mu, \sigma^2)$ distribution truncated between 0 and $\infty$. The hyperparameters listed in stage 3 of (\ref{eqn:hier}) were specifically tuned for our real-data analysis in Section \ref{sec:real}; we detail this tuning process and justify these priors in Section \ref{sec:sim}. In general, these hyperparameters must be carefully tuned and selected, particularly in small-sample settings where prior distributions materially impact inference.    

While it is possible to incorporate dependence into joint prior distributions using copulas (see e.g., \citet{elfadaly2017eliciting, wilson2018specification, hagar2025posterior}), we note that all marginal prior distributions specified in (\ref{eqn:hier}) are independent. By specifying prior distributions on the mean and coefficient of variation, we more effectively capture the relationship between location and dispersion in the marginal distributions compared to specifying independent priors on the means and variances. For example, it is typically implausible for a marginal distribution to exhibit both a very small mean and very large variance. Instead of using truncated normal distributions in (\ref{eqn:hier}), it would be possible to utilise normal distributions for the means, coefficients of variation, and standard deviations on the logarithmic scale. We avoided this specification since it can yield posterior distributions with excessively heavy tails in small-sample settings.

    \subsection{Bayesian Model Averaging}\label{sec:methods.bma}

    In small-sample settings inherent to dose-finding trials, there is limited information to confidently select a joint statistical model comprised of marginal distributions and a copula. The hierarchical model in (\ref{eqn:hier}), however, allows us to weight inference according to a set of candidate models via BMA \citep{hoeting1999bayesian, fragoso2018bayesian}. The set of the candidate models we consider is $\{M_{rst} : r = 1,2;~ s = 1,2;~ t = 1, 2, 3,4 \}$, where $M_{rst}$ is the model defined by marginal distributions $Z_1 = r$ and $Z_2 = s$ and copula $Z_c = t$.

    BMA weights the candidate models based on the observed data. We denote this data -- which consists of the dose assignments and a sample of bivariate responses of size $n = \sum_{j=0}^Jn_j$ across doses -- as $\boldsymbol{y}_n$. Each $M_{rst}$ is assigned a prior probability by way of the Dirichlet priors in stage 3 of (\ref{eqn:hier}). Under BMA, the posterior weight for the candidate model $M_{rst}$ is as follows:
    \begin{equation}\label{eqn:model.prob}
        \Pr(M_{rst}~|~\boldsymbol{y}_n) = \dfrac{\Pr(\boldsymbol{y}_n~|~M_{rst})\Pr(M_{rst})}{\sum_{r'=1}^2\sum_{s'=1}^2\sum_{t'=1}^4 \Pr(\boldsymbol{y}_n~|~M_{r's't'})\Pr(M_{r's't'})},
    \end{equation}
where
\begin{equation*}\label{eqn:marginal.lik}
        \Pr(\boldsymbol{y}_n~|~M_{rst}) = \int \Pr(\boldsymbol{y}_n~|~\boldsymbol{\Delta}_{rst}, M_{rst})\Pr(\boldsymbol{\Delta}_{rst}~|~ M_{rst})d\boldsymbol{\Delta}_{rst}.
    \end{equation*}
    To prompt general notation, we let $\boldsymbol{\Delta}_{rst}$ above represent all random variables in stages 2 and 3 of (\ref{eqn:hier}) under the model $M_{rst}$. 
    
    The hierarchical model in (\ref{eqn:hier}) accommodates a common prior across candidate models via the mean $m_{kj}$, coefficient of variation $cv_{kj}$, and rank correlation $\tau_j$. Even so, it is the marginal distribution parameters, $(\alpha_{kj}, \beta_{kj})$ or $(\mu_{kj}, \sigma_{kj})$, and the copula parameter, $\theta_j$, that are informed by the likelihood component of the posterior as functions of $m_{kj}$, $cv_{kj}$, and $\tau_{j}$. It follows that the posteriors of $m_{kj}$, $cv_{kj}$, and $\tau_{j}$ generally differ across candidate models. These differences do not present issues for inference, as BMA is implicitly implemented by allowing the categorical indicators $(Z_1, Z_2, Z_c)$ to vary within Markov chain Monte Carlo (MCMC) methods. Provided that the MCMC algorithm used to fit (\ref{eqn:hier}) converges to its stationary distribution, valid inference based on BMA is obtained. We provide more details on fitting our hierarchical model and checking Markov chain convergence in Section \ref{sec:code}.

    The posterior probability in (\ref{eqn:model.prob}) can be re-expressed as 
    \begin{equation}\label{eqn:bma.def}
    \Pr(M_{rst}~|~\boldsymbol{y}_n) = \Pr(Z_1 = r, Z_2 = s, Z_c = t~|~\boldsymbol{y}_n). 
    \end{equation}
    Given a sample of $Q$ posterior draws obtained using MCMC, we can estimate this posterior probability as
    \begin{equation*}\label{eqn:bma.est}
       \widehat{\Pr}(M_{rst}~|~\boldsymbol{y}_n) = \dfrac{1}{Q}\sum_{q=1}^Q\mathbb{I}\{z_{1q} =r, z_{2q} = s ,z_{cq} = t \},
    \end{equation*}
where $(z_{1q}, z_{2q}, z_{cq})$ are draws $q = 1, \dots, Q$ from the joint posterior distribution of $Z_1$, $Z_2$, and $Z_c$. We now simplify the notation from (\ref{eqn:hier}) by letting $\boldsymbol{m} = \{\{m_{kj}\}_{j=0}^J\}_{k=1}^2$ and $\boldsymbol{cv} = \{\{cv_{kj}\}_{j=0}^J\}_{k=1}^2$, and $\boldsymbol{\tau} = \{\tau_{j}\}_{j=0}^J$. This simplified notation defines the posterior distribution used for posterior predictive inference in Section \ref{sec:methods.ppp}:
    \begin{equation}\label{eqn:bma.post}
        \Pr(\boldsymbol{m}, \boldsymbol{cv}, \boldsymbol{\tau}~|~\boldsymbol{y}_n) = \sum_{r=1}^2\sum_{s=1}^2\sum_{t=1}^4 \Pr(\boldsymbol{m}, \boldsymbol{cv}, \boldsymbol{\tau}~|~M_{rst}, \boldsymbol{y}_n)\Pr(M_{rst}~|~\boldsymbol{y}_n).
    \end{equation}
This posterior incorporates uncertainty in the candidate models using BMA and contains the parameters required to generate bivariate data from the model in stage 1 of (\ref{eqn:hier}).

    \subsection{Posterior Predictive Inference}\label{sec:methods.ppp}

    We now define the posterior predictive distribution for dose $j = 0, 1, \dots, J$. This posterior predictive distribution characterises the distribution of data from hypothetical future patients receiving dose $j$ according to the posterior distribution in (\ref{eqn:bma.post}). We denote a bivariate observation corresponding to a future patient receiving dose $j$ as $(\tilde{Y}_{1j}, \tilde{Y}_{2j})$. The posterior predictive distribution for dose $j$ is formally defined such that
        \begin{equation}\label{eqn:post.pred}
       (\tilde{Y}_{1j}, \tilde{Y}_{2j}) \sim \iiint\hspace*{-5pt}\iiint \text{BIVAR}(z_{1},m_{1j}, cv_{1j},z_{2},m_{2j}, cv_{2j},  z_{c}, \tau_{j})\Pr(\boldsymbol{m}, \boldsymbol{cv}, \boldsymbol{\tau}~|~\boldsymbol{y}_n)d\boldsymbol{m}d\boldsymbol{cv}d\boldsymbol{\tau}dz_1dz_2dz_c.
    \end{equation}
While we define the posterior predictive distribution using the posterior distribution in (\ref{eqn:bma.post}), only the components of $\boldsymbol{m}$, $\boldsymbol{cv}$, and $\boldsymbol{\tau}$ related to dose $j$ impact (\ref{eqn:post.pred}). Moreover, uncertainty in the model $M_{rst}$ is incorporated into $\Pr(\boldsymbol{m}, \boldsymbol{cv}, \boldsymbol{\tau}~|~\boldsymbol{y}_n)$ by way of the relationships in (\ref{eqn:bma.def}) and (\ref{eqn:bma.post}). 

We aim to conduct inference for dose $j$ using the posterior predictive probability that $\Pr((\tilde{Y}_{1j}, \tilde{Y}_{2j}) \in A~|~\boldsymbol{y}_n)$, where $A$ is some relevant event. These posterior predictive probabilities could then be compared across doses. Suppose each endpoint is defined such that the endpoint increases when the patient has a positive response to treatment. In that case, we may want to calculate the posterior predictive probability for the event $A = \{(\tilde{Y}_{1j}, \tilde{Y}_{2j}) : \tilde{Y}_{1j} > 1, \tilde{Y}_{2j}> 1\}$. Nevertheless, the event $A$ could be defined more broadly. For instance, the arbitrary events $A = \{(\tilde{Y}_{1j}, \tilde{Y}_{2j}) : 1 < \tilde{Y}_{1j} \le 1.1, \tilde{Y}_{2j}< 1\}$ or $A = \{\tilde{Y}_{1j} : \tilde{Y}_{1j} > 0.9\}$ might be relevant based on certain trial objectives. Moreover, dose recommendations could also be based on posterior predictive probabilities for a collection of events. 

Algorithm \ref{alg1} details a procedure to estimate the posterior predictive probability for a single event $A$ across all doses; this algorithm also obtains a $100\times(1-\alpha)\%$ credible interval for $\Pr((\tilde{Y}_{1j}, \tilde{Y}_{2j}) \in A~|~\boldsymbol{y}_n)$ corresponding to each dose. For Algorithm \ref{alg1}, we also must select the number of posterior draws $Q$ and the number of samples $W$ simulated from (\ref{eqn:post.pred}) for each posterior draw $q = 1, \dots, Q$. We recommend using values of at least $Q = 10^4$ and $W = 500$ to obtain reliable inference, and $Q$ should generally be much greater than $W$. We elaborate on several additional aspects of Algorithm \ref{alg1} below.

             \begin{algorithm}
\caption{Procedure to Conduct Posterior Predictive Inference}
\label{alg1}

\begin{algorithmic}[1]
\setstretch{1}
\Procedure{PostPred}{$\boldsymbol{y}_n$, $A$, $J$, $Q$, $W$, $\alpha$}
\State  Use MCMC to obtain posterior draws $\{z_{1q},\boldsymbol{m}_{1q}, \boldsymbol{cv}_{1q}, z_{2q},\boldsymbol{m}_{2q}, \boldsymbol{cv}_{2q}, z_{cq}, \boldsymbol{\tau}_{q}\}_{q=1}^Q$ from (\ref{eqn:hier})
\For{$j$ in $0$:$J$}
\For{$q$ in 1:$Q$}
\For{$w$ in 1:$W$}
\State Simulate $(\tilde{y}_{1jqw}, \tilde{y}_{2jqw}) \sim \text{BIVAR}(\{z_{kq},m_{kjq}, cv_{kjq}\}_{k=1}^2, z_{cq}, \tau_{jq})$ from stage 1 of (\ref{eqn:hier})
\EndFor
\State Compute $\widehat{\Pr}(A_{jq}~|~\boldsymbol{y}_n) = W^{-1}\sum_{w=1}^W \mathbb{I}\{(\tilde{y}_{1jqw}, \tilde{y}_{2jqw}) \in A \}$
\EndFor
\State Compute $\widehat{\Pr}(A_{j}~|~\boldsymbol{y}_n) = Q^{-1}\sum_{q=1}^Q \widehat{\Pr}(A_{jq}~|~\boldsymbol{y}_n)$
\State Construct a $100\times(1-\alpha)\%$ credible interval for $\Pr(A_{j}~|~\boldsymbol{y}_n)$ using quantiles of $\{\widehat{\Pr}(A_{jq}~|~\boldsymbol{y}_n)\}_{q=1}^Q$
\EndFor

 \State \Return $\widehat{\Pr}(A_{j}~|~\boldsymbol{y}_n)$ and the corresponding credible intervals for all doses $j$

\EndProcedure

\end{algorithmic}
\end{algorithm}

In Lines 7 to 10 of Algorithm \ref{alg1}, we use more compact notation for $\Pr((\tilde{Y}_{1j}, \tilde{Y}_{2j}) \in A~|~\boldsymbol{y}_n)$ that incorporates subscripts for the dose $j$ and the posterior draw $q$. To construct a $100\times(1-\alpha)\%$ credible interval for each posterior predictive probability, we take the $\alpha/2$ and $1- \alpha/2$ quantiles of $\{\widehat{\Pr}(A_{jq}~|~\boldsymbol{y}_n)\}_{q=1}^Q$. We consider a single event $A$ in Algorithm \ref{alg1}, but posterior predictive probabilities for a collection of events could easily be obtained by implementing the processes in Lines 7 to 9 for multiple events within the same iteration of each for loop. 

Lastly, we discuss the impact of fitting the model in (\ref{eqn:hier}) to $1/Y_1$ and $Y_2$ on posterior predictive inference. This transformation may be required to ensure positive dependence between the two endpoints. If this transformation is applied, the indicator function in Line 7 should be modified to $\mathbb{I}\{(1/\tilde{y}_{1jqw}, \tilde{y}_{2jqw}) \in A \}$. The remainder of the algorithm can be implemented without modifications. This transformation does not hinder our ability to conduct posterior predictive inference or impact the interpretation of the posterior predictive probabilities. However, the interpretation of the parameters for the marginal distributions and copula for the fitted hierarchical model would not apply to the untransformed data $Y_1$ and $Y_2$.

    \section{Numerical Studies}\label{sec:sim}

    We next detail two sets of numerical studies used to tune the stage-3 prior distributions in (\ref{eqn:hier}). The first set of simulations tuned the priors for the marginal distribution parameters, and the second set informed the priors for the copula parameters. Our methodology proposed in Section \ref{sec:methods} is well suited for small-sample settings inherent to dose-finding trials, and it is crucial to carefully tune prior distributions for small-sample Bayesian analyses before observing data.  

    In the first set of simulations to select priors associated with the marginal distributions, we suppose the data from the two endpoints are independent. We thus consider the four models $\{M_{rs1}: r = 1, 2; ~ s = 1, 2\}$. We henceforth drop the subscript $t=1$ used to denote the independence copula for brevity. To tune the priors across a broad range of data-generating processes, we generate 1000 samples from each of the four $M_{rs}$ models using the hierarchical process in (\ref{eqn:hier}) with the following stage-3 prior distributions: $\mu_{mk} \sim \mathcal{N}(1, 400^{-1})$, $\sigma_{mk} \sim \mathrm{TN}(0.2, 800^{-1}, 0, \infty)$, $\mu_{cvk} \sim \mathcal{N}(0.5, 400^{-1})$, $\sigma_{cvk} \sim \mathrm{TN}(0.1, 800^{-1}, 0, \infty)$ for $k = 1, 2$. We consider $J=3$ active doses and a placebo for the numerical studies in this section. All generated samples had the same dose composition as DEN-181 described in Section \ref{sec:den} (i.e., $n = 16$ across all doses with $n_0 = 5$, $n_1 = 4$, $n_2 = 3$, and $n_3 = 4$). 
    
    The prior distributions used for data generation have the same location parameters as those used for data analysis in (\ref{eqn:hier}), but the data-generating priors have much less dispersion. We note that small changes in $\mu_{mk}$, $\sigma_{mk}$, $\mu_{cvk}$, and $\sigma_{cvk}$ can collectively result in large changes to the data-generating process since additional variability is propagated into stages 2 and 1. In general, these prior distributions reflect a scenario where we expect (i) the typical dose (including the placebo) to have a weak effect, (ii) a moderate level of dispersion in the marginal distribution for each dose, and (iii) a moderate level of variability in the marginal distributions across doses.

    The prior distributions used for data analysis in (\ref{eqn:hier}) that are related to the copula are not relevant for this first set of simulations since we enforce the constraint that $Z_c = 1$. Our primary aim with these first simulations is to select parameters for the Dirichlet priors on $\pi_1$ and $\pi_2$ in stage 3 of (\ref{eqn:hier}). If the true data-generating model is $M_{rs}$, we generally want $\widehat{Pr}(M_{rs}~|~\boldsymbol{y}_n)$ to be greater, on average, than the posterior probabilities for the other models. In initial simulations using $\mathrm{Dir}(1, 1)$ priors for both $\pi_1$ and $\pi_2$, we observed bias toward lognormal marginals. Even when the data were generated according to $M_{11}$, $M_{12}$, or $M_{21}$, $\widehat{Pr}(M_{22}~|~\boldsymbol{y}_n)$ was often the largest posterior probability. This bias occurs in small-sample settings because, even though a common prior is specified on the means and coefficients of variation across models, different prior distributions are \emph{induced} on the model-specific gamma or lognormal parameters. All posterior distributions in this section were approximated using MCMC with 16 chains, where the first 1000 iterations from each chain were discarded as burn-in and the following 2500 posterior draws were retained.

    To obtain better marginal distribution selection in our small-sample setting, we ran further simulations where we increased the value of the first parameter in the two-parameter Dirichlet prior. Increasing this parameter gives higher prior probabilities to models $M_{rs}$ containing a gamma marginal distribution. The smallest value for this parameter that resulted in well calibrated model selection was 1.625, as specified in (\ref{eqn:hier}). The $\mathrm{Dir}(1.625, 1)$ prior is such that the prior probabilities for the gamma and lognormal marginals are $0.619$ and $0.381$, respectively.

\begin{figure}[!tb]
      \centering
		\includegraphics[width = 0.7\textwidth]{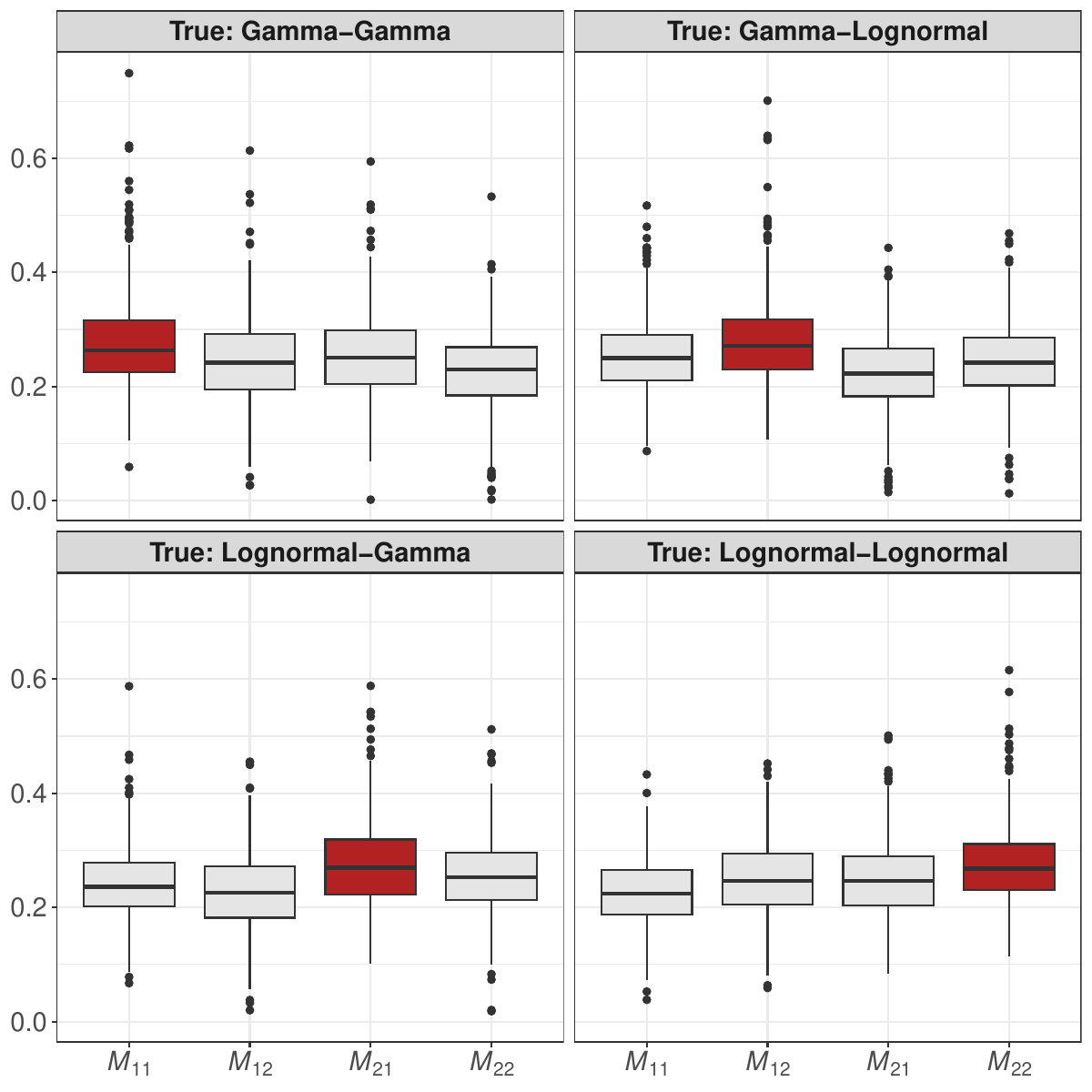} 

		\caption{\label{fig:marg} Boxplots of 1000 estimated posterior probabilities ${Pr}(M_{rs}~|~\boldsymbol{y}_n)$ for each data-generating model. The red boxplots correspond to the correct marginal distributions.} 
	\end{figure}

    When this Dirichlet prior is used for data analysis, Figure \ref{fig:marg} visualises the estimated posterior probabilities corresponding to the 1000 simulated samples from each model $M_{rs}$. For each data-generating process, the boxplot of posterior probabilities for the correct model $M_{rs}$ (in red) is slightly higher than those for the incorrect models (in grey). Table \ref{tab:marg} details the proportion of simulation repetitions in which the estimated posterior probability is largest for each data-generating and candidate model combination. Both Figure \ref{fig:marg} and Table \ref{tab:marg} confirm that the posterior probabilities are well calibrated, but the ability to discriminate between candidate models is limited due to the small-sample setting. 

    \begin{table}[!htb]
\centering
\begin{tabular}{c|cccc}
   True Marginals         & $M_{11}$ & $M_{12}$ & $M_{21}$ & $M_{22}$ \\ \hline
Gamma-Gamma    & 0.336   & 0.240     & 0.250      & 0.174       \\
Gamma-Lognormal   & 0.234   & 0.370     & 0.178     & 0.218       \\
Lognormal-Gamma   & 0.174   & 0.180     & 0.388     & 0.258       \\
Lognormal-Lognormal & 0.170    & 0.264     & 0.236     & 0.330      
\end{tabular}
\caption{Proportion of 1000 simulation repetitions in which $\widehat{Pr}(M_{rs}~|~\boldsymbol{y}_n)$ is the largest for each candidate model (column) and true data-generating model (row). The correct marginal distributions are along the diagonal. }\label{tab:marg}
\end{table}

Nevertheless, this performance is suitable for our proposed method. We aim to estimate posterior predictive probabilities related to joint events for the two endpoints, so it is of greater importance to discriminate between dependence structures. We simply accommodate uncertainty in the marginals so that posterior predictive inference is not predicated on a single choice for these distributions. In Appendix A of the online supplement, we repeat this numerical study for the marginal distributions with samples of size $n = 160$ and a $\text{Dir}(1, 1)$ prior for $\pi_1$ and $\pi_2$. For that large-sample study, better discrimination between the models $M_{rs}$ is observed, and the Dirichlet priors have less impact on posterior inference.

In the second set of simulations to select priors related to the copula, we suppose both marginal distributions are gamma. We now consider the four models $\{M_{11t}: t = 1, 2, 3, 4\}$. We henceforth drop the subscripts $r=1$ and $s=1$ used to denote the gamma marginals for brevity. To implement data generation, we modified the hierarchal process for the first set of simulations. The sample sizes from DEN-181 and the stage-3 priors on $\mu_{mk}$, $\sigma_{mk}$, $\mu_{cvk}$, and $\sigma_{cvk}$ were retained. In addition, the priors $\mu_\tau \sim \mathcal{N}(0.25, 20^{-1})$ and $\sigma_\tau \sim \mathrm{TN}(0.25, 20^{-1}, 0, \infty)$ were also used. The data-generating priors have the same location parameters but less dispersion than those used for data analysis in (\ref{eqn:hier}). These priors for the copula parameter ensure that Kendall's $\tau$ is greater than 0.5 in approximately 75\% of the 1000 samples simulated from $M_2$, $M_3$, and $M_4$. We consider samples with relatively strong dependence in this simulation because it would not be a substantial error for $\widehat{Pr}(M_1~|~\boldsymbol{y}_n)$ corresponding to the independence copula to be large if the true underlying dependence is weak. It is not an issue to assume moderate dependence for $M_2$, $M_3$, and $M_4$ a priori since the independence copula is incorporated into the candidate model $M_1$.  

\begin{figure}[!tb]
      \centering
		\includegraphics[width = 0.7\textwidth]{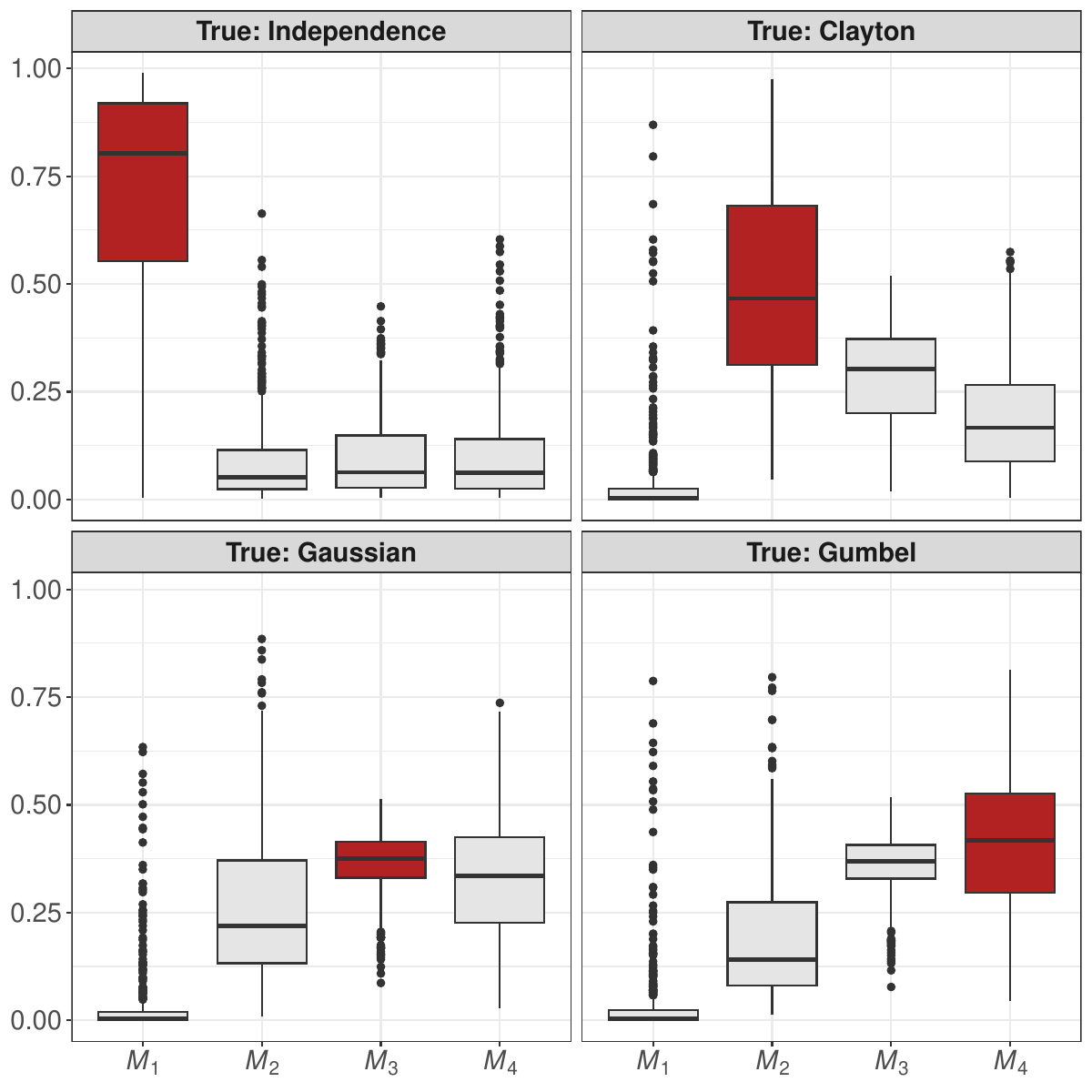} 

		\caption{\label{fig:cop.sim} Boxplots of 1000 estimated posterior probabilities ${Pr}(M_{t}~|~\boldsymbol{y}_n)$ for each data-generating model. The red boxplots correspond to the correct copula.} 
	\end{figure}

Our objective for this second set of simulations is to select parameters for the Dirichlet prior on $\pi_c$ in stage 3 of (\ref{eqn:hier}). When a $\text{Dir}(1, 1, 1.05, 1)$ prior is used for data analysis, Figure \ref{fig:cop.sim} visualises the estimated posterior probabilities corresponding to the 1000 simulated samples from each model $M_{t}$. For each data-generating process, the estimated posterior probabilities for the correct model $M_{t}$ (in red) are generally largest. Table \ref{tab:cop.sim} details the proportion of simulation repetitions in which the estimated posterior probability is largest for each data-generating and candidate copula combination.

\begin{table}[!htb]
\centering
\begin{tabular}{c|cccc}
   True Copula         & $M_{1}$ & $M_{2}$ & $M_{3}$ & $M_{4}$ \\ \hline
Independence & 0.872    & 0.042     & 0.040     & 0.046        \\
Clayton      & 0.030   & 0.638     & 0.272     & 0.060        \\
Gaussian     & 0.030   & 0.240     & 0.442     & 0.288       \\
Gumbel       & 0.032   & 0.128     & 0.346     & 0.494          
\end{tabular}
\caption{Proportion of 1000 simulation repetitions in which $\widehat{Pr}(M_{t}~|~\boldsymbol{y}_n)$ is the largest for each candidate model (column) and true data-generating model (row). The correct copula is along the diagonal. }\label{tab:cop.sim}
\end{table}

Both Figure \ref{fig:cop.sim} and Table \ref{tab:cop.sim} confirm that the posterior probabilities are well calibrated. This calibration indicates that the $\text{Dir}(1, 1, 1.05, 1)$ prior for $\pi_c$ is suitable for data analysis; this Dirichlet distribution assigns a prior probability of 0.259 to the Gaussian copula and a prior probability of 0.247 to each of the three other copulas. Even within this small-sample setting, we are able to reasonably discriminate between dependence structures, which is not surprising given that the non-independence copulas substantially differ when Kendall's $\tau$ is large. In initial simulations using a $\mathrm{Dir}(1, 1, 1, 1)$ prior for $\pi_c$, we observed a very slight bias toward the Gumbel copula when generating data under $M_3$ that was corrected when increasing the third parameter of the Dirichlet distribution to 1.05. Figure \ref{fig:cop.sim} demonstrates that it may be challenging to distinguish between (i) the Clayton and Gaussian copulas or (ii) the Gaussian and Gumbel copulas in small-sample settings; this performance is expected given the coarse similarities between these pairs of copulas in Figure \ref{fig:cop}.  In Appendix A, we repeat this numerical study with samples of size $n = 160$ and a $\text{Dir}(1, 1, 1, 1)$ prior for $\pi_c$, and we observe a greater ability to distinguish between copulas. We use the prior distributions tuned using simulation in this section to implement our real-data analysis in Section \ref{sec:real}.

    \section{Real-Data Analysis}\label{sec:real}

    We now apply COBRA-DOSE to the data from DEN-181 described in Section \ref{sec:den} for illustration. Both biomarkers (the total T cell count as $k = 1$ and the inflammatory dendritic cell count as $k = 2$) were measured at Days 1 and 8. We now summarise each endpoint for patient $i$ receiving dose $j$ as the ratio of the Day 8 value to the Day 1 value (denoted $y_{kji}$), to reflect the changes associated with early immunomodulatory activity. %In the rare event of missing biomarker data, the missing value was imputed using simple linear regression fitted to the available measurements for that patient, with the imputed value obtained from the resulting patient-specific intercept and slope.

    The two biomarkers exhibit negative dependence, so we fit the hierarchical model in (\ref{eqn:hier}) to $1/Y_1$ and $Y_2$. We implemented MCMC with 16 chains, with one chain initialised within each model $M_{rst}$. The first $5 \times 10^3$ iterations from each chain were discarded as burn-in, and every other posterior draw from the remaining $2 \times 10^4$ draws was retained (i.e., the level of thinning was 2). This process resulted in $10^4$ posterior draws per chain, and a sample from the posterior of size $Q = 1.6\times10^5$. Suitable convergence of the Markov chains was verified using $\hat{R}$ statistics \citep{gelman1992inference}; for all variables in Line 2 of Algorithm \ref{alg1}, the maximum $\hat{R}$ value was less than 1.004. The corresponding trace plots also supported the suitable Markov chain convergence.

    Table \ref{tab:den.model} details the estimated posterior probabilities for each of the 16 candidate models given the real data. While there is substantial uncertainty in the best-fitting marginal distributions due to the limited sample size, the gamma distribution is slightly more likely for both the first $(Z_1 = 1)$ and second $(Z_2 = 1)$ endpoints. Of the copula families considered in our analysis, the Gaussian $(Z_c = 3)$ and Gumbel $(Z_c = 4)$ copulas appear to provide the best fit for the data. Given that the sum of the posterior probabilities corresponding to all candidate models with the independence copula is only 0.002, there is value in accounting for the dependence structure in our joint modelling approach.

            \begin{table}[!tb]
    \centering
\begin{tabular}{c|ccccc}
    & $Z_1 = 1, Z_2 = 1$ & $Z_1 = 1, Z_2 = 2$ & $Z_1 = 2, Z_2 = 1$ & $Z_1 = 2, Z_2 = 2$ & All    \\ \hline
$Z_c = 1$ & 0.001  & 0.001  & 0.000  & 0.000  & 0.002 \\
$Z_c = 2$ & 0.017  & 0.015  & 0.011  & 0.009  & 0.052 \\
$Z_c = 3$ & 0.141  & 0.123  & 0.086  & 0.077  & 0.427 \\
$Z_c = 4$ & 0.168  & 0.147  & 0.109  & 0.095  & 0.519  \\
   All     & 0.327  & 0.286  & 0.206  & 0.181  &     
\end{tabular}
\caption{Estimated posterior probability of each candidate model.  }\label{tab:den.model}
\end{table}

We next incorporate this uncertainty over the marginal distributions and copulas when estimating posterior predictive probabilities using Algorithm \ref{alg1} with $W = 500$ samples per posterior draw. For our analysis, a positive response to treatment is characterised by an increase in the total T cell count and a decrease in the inflammatory dendritic cell count. We thus consider the following four events $A$ for posterior predictive inference: $\{(\tilde{Y}_{1j}, \tilde{Y}_{2j}) : Y_{1j} > 1, Y_{2j} < 1\}$, $\{(\tilde{Y}_{1j}, \tilde{Y}_{2j}) : Y_{1j} > 1, Y_{2j} \ge 1\}$, $\{(\tilde{Y}_{1j}, \tilde{Y}_{2j}) : Y_{1j} \le 1, Y_{2j} < 1\}$, and $\{(\tilde{Y}_{1j}, \tilde{Y}_{2j}) : Y_{1j} \le 1, Y_{2j} \ge 1\}$. These events respectively define settings where dose $j$ (i) positively impacts both endpoints, (ii) positively impacts endpoint 1 but negatively impacts endpoint 2, (iii) positively impacts endpoint 2 but negatively impacts endpoint 1, and (iv) negatively impacts both endpoints.  

Table \ref{tab:den.bivar} details the estimated posterior predictive probabilities and associated 95\% credible intervals for each event $A$ described above across all doses. Dose 1 clearly has the largest posterior predictive probability of positively impacting both endpoints as well as the smallest posterior predicative probability of negatively impacting both endpoints. Although there is some overlap in the credible intervals for the posterior predictive probabilities across doses due to the small-sample setting, we emphasise that the probabilities in Table \ref{tab:den.bivar} provide clinicians with intuitive and interpretable data summaries for dose recommendation. The resulting inference is not predicated on the correctness of a single joint distribution for the bivariate data. Furthermore, our method circumvents the cognitive burden on clinicians associated with defining a potentially complex utility function to combine information and resolve trade-offs between endpoints. Instead, posterior predicative probabilities for various joint events can transparently be reported.

%Although the 95\% credible intervals are wide due to the small sample size, the point estimates in the first and last columns for dose 1 are not contained in the corresponding credible intervals for the placebo (dose 0). 

    \begin{table}[!tb]
    \centering
\begin{tabular}{c|cccc}
Dose & $\tilde{Y}_{1j} > 1, \tilde{Y}_{2j} < 1$              & $\tilde{Y}_{1j} > 1, \tilde{Y}_{2j} \ge 1$              & $\tilde{Y}_{1j} \le 1, \tilde{Y}_{2j} < 1$              & $\tilde{Y}_{1j} \le 1, \tilde{Y}_{2j} \ge 1$              \\ \hline
0 & 0.391 (0.177, 0.609) & 0.215 (0.075, 0.407) & 0.047 (0.003, 0.158) & 0.346 (0.165, 0.547) \\
1 & 0.565 (0.271, 0.859) & 0.059 (0.001, 0.243) & 0.152 (0.031, 0.354) & 0.224 (0.028, 0.502) \\
2 & 0.478 (0.206, 0.738) & 0.113 (0.007, 0.341) & 0.118 (0.016, 0.317) & 0.290 (0.072, 0.547) \\
3 & 0.324 (0.053, 0.645) & 0.442 (0.175, 0.768) & 0.009 (0.000, 0.068) & 0.225 (0.030, 0.524)
\end{tabular}
\caption{$\widehat{Pr}((\tilde{Y}_{1j}, \tilde{Y}_{2j}) \in A~|~\boldsymbol{y}_n)$ and 95\% credible intervals for for various events $A$ (columns) and doses $j$ (rows).  }\label{tab:den.bivar}
\end{table}

To illustrate the value of COBRA-DOSE, we fit a simplified version of the model in (\ref{eqn:hier}) to the same data. In this simplified model, each endpoint is considered separately. We thus accommodate uncertainty in the marginal distribution for each endpoint, but copulas are not used to model the dependence structure (i.e., the independence copula is effectively assumed). Table \ref{tab:den.uni} compiles the estimated posterior predictive probabilities and associated 95\% credible intervals for the events $A$ corresponding to positive responses for each endpoint. In this univariate analysis, it is difficult to compare dose 3 with the placebo, for instance, because dose 3 appears more beneficial for endpoint 1 but less promising for endpoint 2. Based on the first and last columns of Table \ref{tab:den.bivar} populated under our joint modelling approach, dose 3 is 6.7\% less likely to have a positive impact on both endpoints but 12.1\% more likely to not negatively impact both endpoints, providing practitioners with this additional context.

    \begin{table}[!tb]
    \centering
\begin{tabular}{c|cc}
Dose & $\tilde{Y}_{1j} > 1$              & $\tilde{Y}_{2j} < 1$               \\ \hline
0 & 0.594 (0.370, 0.792) & 0.457 (0.182, 0.716) \\
1 & 0.667 (0.376, 0.916) & 0.684 (0.377, 0.969) \\
2 & 0.622 (0.339, 0.873) & 0.584 (0.227, 0.934) \\
3 & 0.734 (0.398, 0.970) & 0.347 (0.050, 0.678)
\end{tabular}
\caption{$\widehat{Pr}(\tilde{Y}_{kj} \in A~|~\boldsymbol{y}_n)$ and 95\% credible intervals for unvariate events $A$ that characterise positive treatment response (columns) and doses $j$ (rows).  }\label{tab:den.uni}
\end{table}

Under an independence assumption, approximations to the joint posterior predictive probabilities in Table \ref{tab:den.bivar} could be obtained by multiplying the marginal posterior predictive probabilities in the two columns of Table \ref{tab:den.uni}. For dose 2, the resulting approximation to the probability in the first column of Table \ref{tab:den.bivar} is 0.363; the estimated posterior predictive probability that accounts for the dependence structure is 0.478 (corresponding to a relative increase of 31.7\% over 0.363). It is therefore important to account for the dependence structure when conducting inference based on posterior predictive probabilities related to \emph{joint} events.

\begin{figure}[!tb]
      \centering
		\includegraphics[width = 0.9\textwidth]{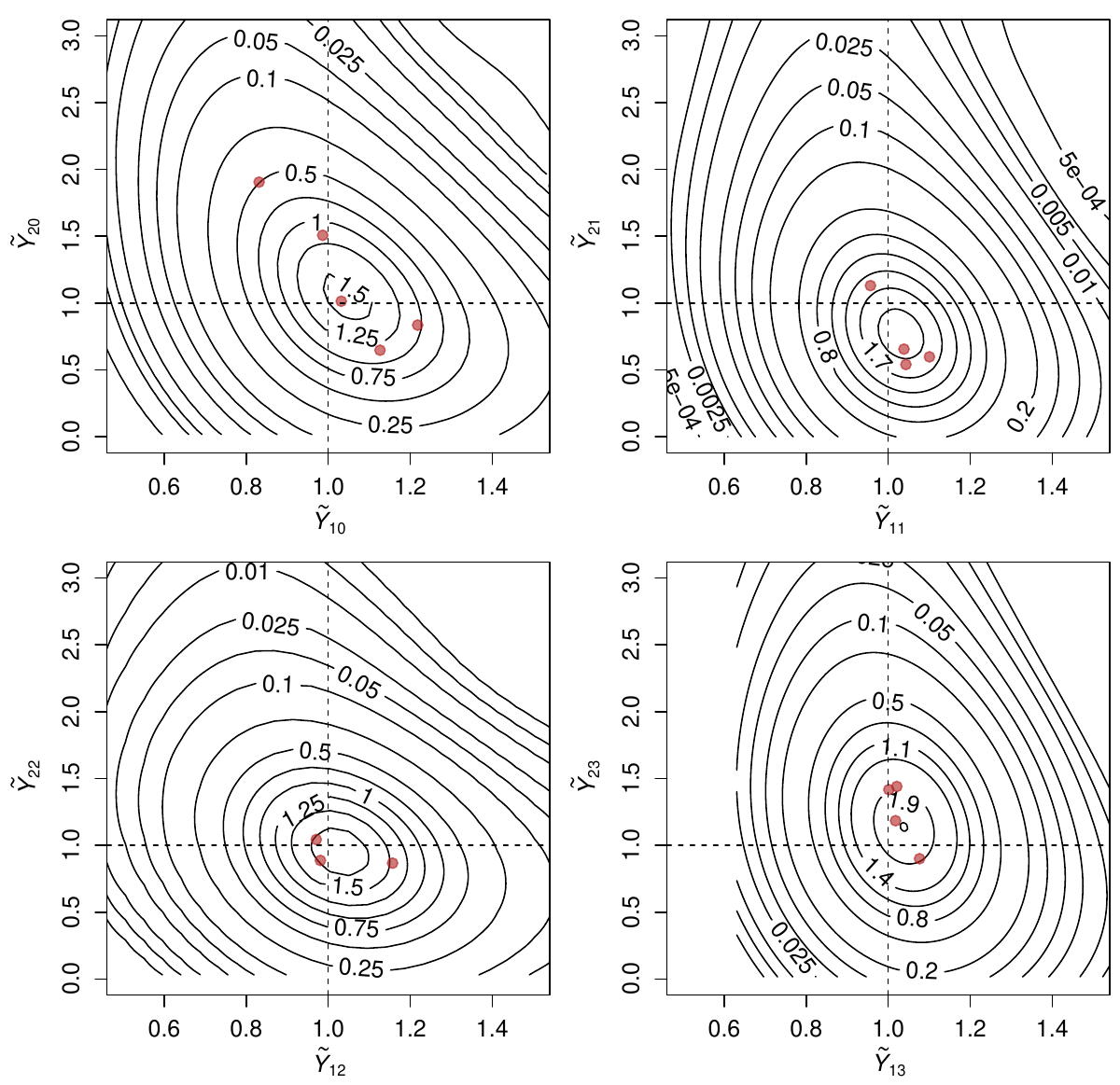} 

		\caption{\label{fig:contour} Contour plots of the bivariate posterior predictive distribution of $(\tilde{Y}_{1j}, \tilde{Y}_{2j})~|~\boldsymbol{y}_n$ for doses $j = 0, 1, 2, 3$. The red points correspond to the observed data from dose $j$. The dashed lines at $\tilde{Y}_{1j} = 1$ and $\tilde{Y}_{2j} = 1$ distinguish between positive and negative responses for each endpoint.} 
	\end{figure}

We note that our proposed approach does rely on the assumption that the hierarchical model in (\ref{eqn:hier}) is reasonable. This assumption can be verified by plotting the posterior predictive distribution for dose $j$ in (\ref{eqn:post.pred}) along with the observed data. Figure \ref{fig:contour} visualises the posterior predictive distribution and observed data across all doses using contour plots. These contour plots illustrate a similar structure between the posterior predictive distributions across doses prompted by the hierarchical modelling approach. Nevertheless, the location and dispersion of these bivariate distributions differ across doses. For each dose, the red points corresponding to the observed data in $\boldsymbol{y}_n$ appear to be situated in regions of the contour plot with relatively high density. If the patterns in the observed data are vastly different across doses or the contour plots of the posterior predictive distributions do not provide a good fit to the data, a hierarchical approach to joint modelling may not be appropriate, and alternative modelling approaches should be considered. For the real data, the model in (\ref{eqn:hier}) provides a decent fit, corroborating the analysis in this section.  

    \section{Software}\label{sec:code}

    We use the \texttt{nimble} package in R \citep{de2017programming, de2026nimble} to flexibly implement MCMC for the hierarchical model in (\ref{eqn:hier}). The \texttt{nimble} package allows us to define and compile custom PDFs that characterise the joint behaviour of multiple endpoints using copulas. Standard random walk samplers were used for the variables in (\ref{eqn:hier}) that have support over $\mathbb{R}$, slice samplers \citep{neal2003slice} were employed for the variables that are strictly positive, conjugate samplers were leveraged for $\pi_1$, $\pi_2$, and $\pi_c$, and categorical samplers were used for $Z_1$, $Z_2$, and $Z_c$. 

    To facilitate the implementation of the methods proposed in this paper, we have developed the \texttt{CobraDose} package in R \citep{hagar2026cobra}. For a general example, this R package can be used to (i) sample from the posterior distribution of the model in (\ref{eqn:hier}), (ii) check the convergence of resulting Markov chains, and (iii) estimate posterior predictive probabilities and their associated credible intervals using Algorithm \ref{alg1}. We also provide functions that write R scripts to tune the priors (in parallel across multiple cores) for the parameters associated with the marginal distributions and copulas as in Section \ref{sec:sim}. Further details are provided in the user manual for the \texttt{CobraDose} package. 

    \section{Discussion}\label{sec:disc}

    In this paper, we proposed an approach to help select the optimal biological dose in early-phase dose-selection trials with two endpoints. Our proposed method provides clinicians with interpretable posterior predictive probabilities concerning events related to various combinations of outcomes across doses. These posterior predictive probabilities are based on a posterior distribution that accommodates uncertainty in the marginal distributions and copula via Bayesian model averaging. The resulting dose recommendations are thus not predicated on a single model. Moreover, we need not use a potentially complex utility function to synthesise information from multiple endpoints into a single measure since posterior predictive probabilities for various joint events can transparently be reported. While our methodology is illustrated using the DEN-181 trial, our proposed approach can be applied to any dose-selection analysis based on two continuous and positive biomarkers. Our method can be readily implemented using the developed \texttt{CobraDose} package in R.

    Future research could extend the methodology from this paper in several aspects. First and foremost, our framework could be adapted to accommodate more than two endpoints, which would be useful in an exploratory trial with limited knowledge of which biomarkers are related to the response. Archimedean copulas, including the Clayton and Gumbel copulas, parsimoniously generalise to model dependence in more than two dimensions, whereas vine copulas \citep{joe1996families, bedford2002vines} offer a more complex and flexible solution to model multivariate dependence. Furthermore, the scope of our proposed approach could be broadened to accommodate binary or count endpoints for which it may not be sensible to consider relative changes as a continuous variable. For such contexts, copulas might be used to model dependence between continuous endpoints and latent variables related to non-continuous outcomes. Lastly, we used standard samplers to implement MCMC in this paper, but it may be possible to develop custom samplers that improve the computational efficiency of posterior approximation for the small-sample, hierarchical models on which we base inference.

 \section*{Supplementary Material}
These materials include additional numerical studies from Section \ref{sec:sim}. An R package to implement our proposed approach is available online: \url{https://github.com/UQ-ULTRA/CobraDose}.

% \section*{Acknowledgements}

	\section*{Funding}
	%%%%%%%%%%%%%%%%%%%%%%%%%%%%%%%%%%%%%%%%%%%%%%%%%%%%%%%%%%%%%%%%%%%%%%%%%%%%%%%%%%%%%%%%%%%%%%%%%%%%%
	%%%%%%%%%%%%%%%%%%%%%%%%%%%%%%%%%%%%%%%%%%%%%%%%%%%%%%%%%%%%%%%%%%%%%%%%%%%%%%%%%%%%%%%%%%%%%%%%%%%%%
 
This project was supported by funding from the Australian Trials Methodology (AusTriM) Research Network, a Centre of Research Excellence grant from the National Health and Medical Research Council, ID\# 1171422.
	
		%%%%%%%%%%%%%%%%%%%%%%%%%%%%%%%%%%%%%%%%%%%%%%%%%%%%%%%%%%%%%%%%%%%%%%%%%%%%%%%%%%%%%%%%%%%%%%%%%%%%%
	%%%%%%%%%%%%%%%%%%%%%%%%%%%%%%%%%%%%%%%%%%%%%%%%%%%%%%%%%%%%%%%%%%%%%%%%%%%%%%%%%%%%%%%%%%%%%%%%%%%%%
% 	\appendix
% \numberwithin{equation}{section}
% \renewcommand{\theequation}{\thesection.\arabic{equation}}

% \numberwithin{figure}{section}
% \renewcommand{\thefigure}{\thesection.\arabic{figure}}

\bibliographystyle{chicago}

% \bibliography{cobra}

\end{document}

\newcommand{\bb}{\boldsymbol{\beta}}

	\title{COBRA-DOSE: Copula-based Bayesian Model Averaging for Dose Selection\medskip\\
 \Large{Supplementary Material}}

 	\author{Luke Hagar \hspace{35pt} Min Zhang$^*$ \hspace{35pt} Ranjeny Thomas$^{\dagger}$ \hspace{35pt} Andrew J. Martin$^*$ \bigskip \\ 
 $^*$\textit{Clinical Trials Capability, The University of Queensland} \\ $^{\dagger}$\textit{Frazer Institute, The University of Queensland}}

	\date{}

	\maketitle

	\baselineskip=19.5pt

% 	%%%%%%%%%%%%%%%%%%%%%%%%%%%%%%%%%%%%%%%%%%%%%%%%%%%%%%%%%%%%%%%%%%%%%%%%%%%%%%%%%%%%%%%%%%%%%%%%%%%%%
% 	%%%%%%%%%%%%%%%%%%%%%%%%%%%%%%%%%%%%%%%%%%%%%%%%%%%%%%%%%%%%%%%%%%%%%%%%%%%%%%%%%%

% 	%%%%%%%%%%%%%%%%%%%%%%%%%%%%%%%%%%%%%%%%%%%%%%%%%%%%%%%%%%%%%%%%%%%%%%%%%%%%%%%%%%%%%%%%%%%%%%%%%%%%%
% 	%%%%%%%%%%%%%%%%%%%%%%%%%%%%%%%%%%%%%%%%%%%%%%%%%%%%%%%%%%%%%%%%%%%%%%%%%%%%%%%%%%%%%%%%%%%%%%%%%%%%%

		%%%%%%%%%%%%%%%%%%%%%%%%%%%%%%%%%%%%%%%%%%%%%%%%%%%%%%%%%%%%%%%%%%%%%%%%%%%%%%%%%%%%%%%%%%%%%%%%%%%%%
	%%%%%%%%%%%%%%%%%%%%%%%%%%%%%%%%%%%%%%%%%%%%%%%%%%%%%%%%%%%%%%%%%%%%%%%%%%%%%%%%%%%%%%%%%%%%%%%%%%%%%

	\appendix
\numberwithin{equation}{section}
\renewcommand{\theequation}{\thesection.\arabic{equation}}

\numberwithin{figure}{section}
\renewcommand{\thefigure}{\thesection.\arabic{figure}}

\numberwithin{table}{section}
\renewcommand{\thetable}{\thesection.\arabic{table}}

\numberwithin{algorithm}{section}
\renewcommand{\thealgorithm}{\thesection.\arabic{algorithm}}

 \section{Additional Simulations for Section 4}\label{sec:large}

 We now detail the results from the additional numerical studies with larger samples. In these simulations, we generated 100 samples of size $n = 160$ (the composition across the $J=3$ active doses and placebo for each sample was $n_0 = 50$, $n_1 = 40$, $n_2 = 30$, and $n_3 = 40$). For the first set of simulations related to the marginal distributions, all MCMC settings, parameters for data generation, and priors used for data analysis were the same except the $\text{Dir}(1,1)$ priors now used for $\pi_1$ and $\pi_2$ from (3) of the main text.  
 
\begin{figure}[!htb]
      \centering
		\includegraphics[width = 0.7\textwidth]{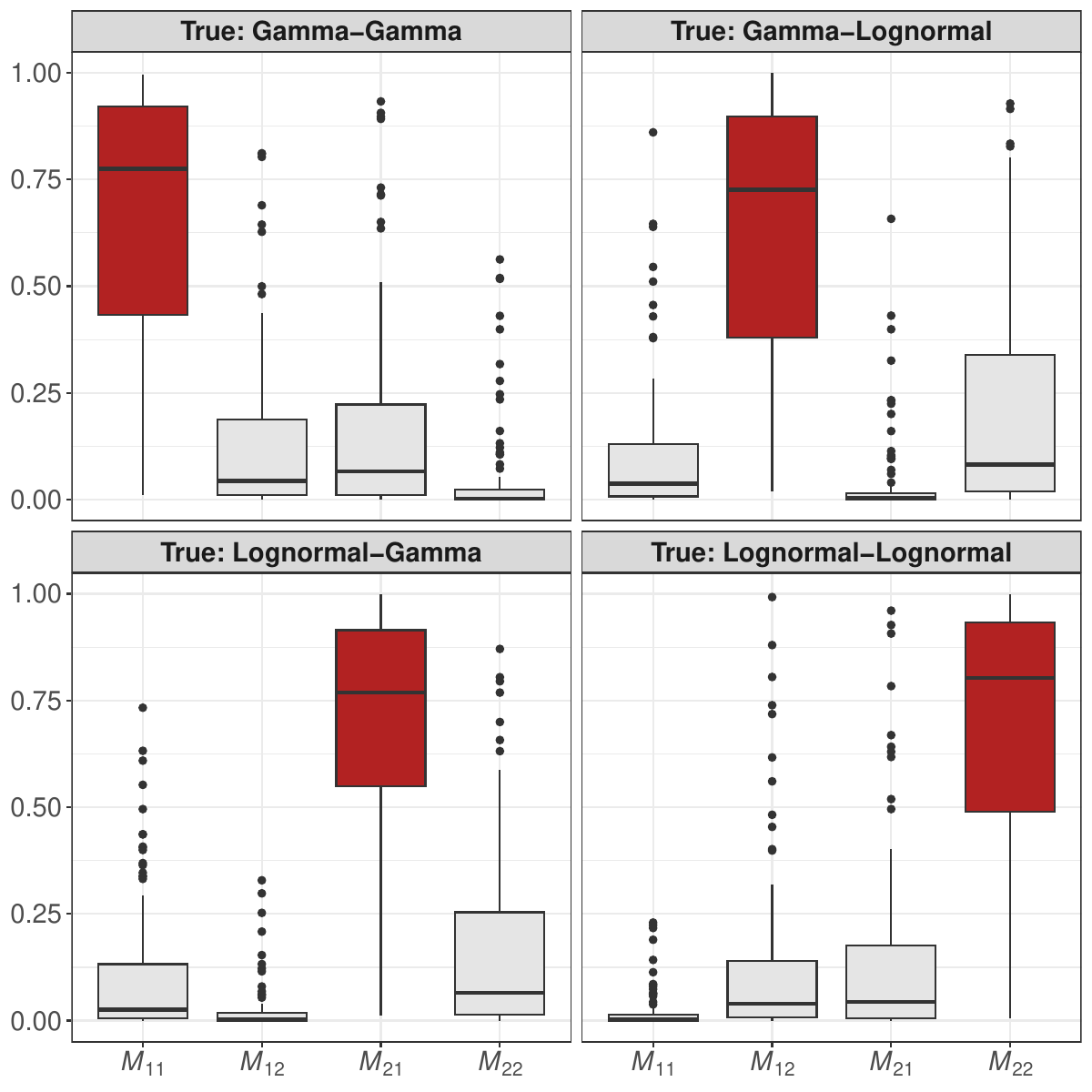} 

		\caption{\label{fig:marg.large} Boxplots of 100 estimated posterior probabilities ${Pr}(M_{st}~|~\boldsymbol{y}_n)$ for each data-generating model with larger samples. The red boxplots correspond to the correct marginal distributions.} 
	\end{figure}

 Figure \ref{fig:marg.large} visualises the estimated posterior probabilities corresponding to the 100 simulated samples from each model $M_{rs}$. For each data-generating process, the boxplot of posterior probabilities for the correct model $M_{rs}$ (in red) is now much higher than those for the incorrect models (in grey). Table \ref{tab:marg.large} details the proportion of simulation repetitions in which the estimated posterior probability is largest for each data-generating and candidate model combination. 

\begin{table}[!htb]
\centering
\begin{tabular}{c|cccc}
   True Marginals         & $M_{11}$ & $M_{12}$ & $M_{21}$ & $M_{22}$ \\ \hline
Gamma-Gamma    & 0.75 &	0.09 &	0.12 & 0.04 \\
Gamma-Lognormal   & 0.06 & 0.69 & 0.03 & 0.22       \\
Lognormal-Gamma   & 0.09    & 0.00     & 0.78     & 0.13       \\
Lognormal-Lognormal & 0.00 & 0.09 & 0.12 & 0.81      
\end{tabular}
\caption{Proportion of 100 simulation repetitions with larger samples in which $\widehat{Pr}(M_{rs}~|~\boldsymbol{y}_n)$ is the largest for each candidate model (column) and true data-generating model (row). The correct marginal distributions are along the diagonal. }\label{tab:marg.large}
\end{table}

Both Figure \ref{fig:marg.large} and Table \ref{tab:marg.large} confirm that the posterior probabilities are well calibrated, with improved ability to discriminate between candidate models in this large-sample setting. With the $\text{Dir}(1,1)$ priors for $\pi_1$ and $\pi_2$, there is still a slight bias toward the lognormal distribution, but this bias is much less pronounced than for the small-sample analyses in the main paper. As expected, tuning the prior distributions is therefore less important for a large-sample analysis.

\begin{figure}[!tb]
      \centering
		\includegraphics[width = 0.7\textwidth]{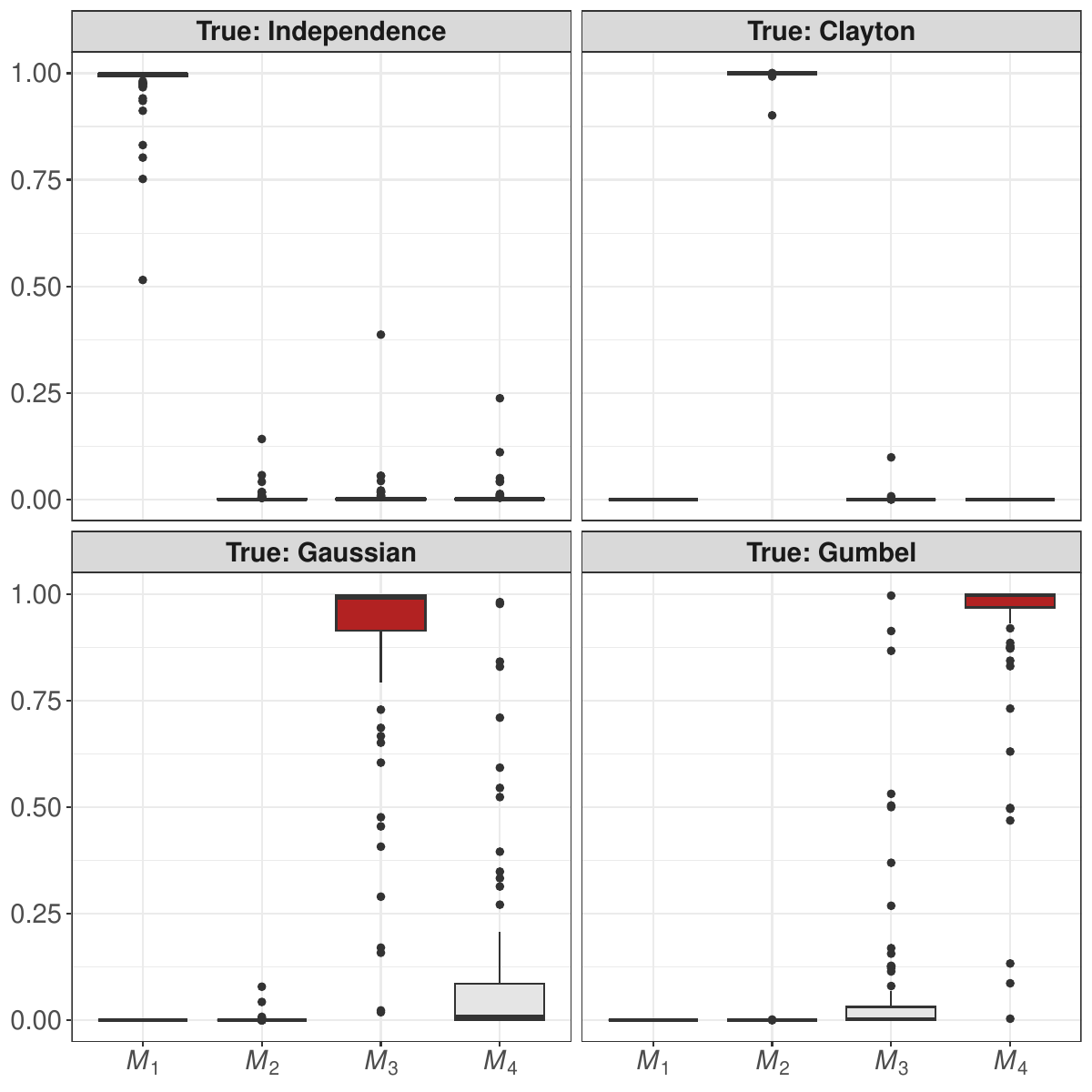} 

		\caption{\label{fig:cop.large} Boxplots of 100 estimated posterior probabilities ${Pr}(M_{t}~|~\boldsymbol{y}_n)$ for each data-generating model with larger samples. The red boxplots correspond to the correct copula.} 
	\end{figure}

For the second set of simulations related to the copulas, all MCMC settings, parameters for data generation, and priors used for data analysis were the same except the $\text{Dir}(1,1, 1, 1)$ prior now used for $\pi_c$ from (3) of the main text. Figure \ref{fig:cop.large} visualises the estimated posterior probabilities corresponding to the 100 simulated samples from each model $M_{t}$. For each data-generating process, the estimated posterior probabilities for the correct model $M_{t}$ (in red) are generally close to 1, except for a few simulation repetitions where there are challenges distinguishing between the Gaussian and Gumbel copulas. 

\begin{table}[!b]
\centering
\begin{tabular}{c|cccc}
   True Copula         & $M_{1}$ & $M_{2}$ & $M_{3}$ & $M_{4}$ \\ \hline
Independence & 1.00 & 0.00 & 0.00 & 0.00        \\
Clayton      & 0.00   & 1.00     & 0.00    & 0.00        \\
Gaussian     & 0.00   & 0.00     & 0.92     & 0.08       \\
Gumbel       & 0.00   & 0.00     & 0.06     & 0.94          
\end{tabular}
\caption{Proportion of 100 simulation repetitions with larger samples in which $\widehat{Pr}(M_{t}~|~\boldsymbol{y}_n)$ is the largest for each candidate model (column) and true data-generating model (row). The correct copula is along the diagonal. }\label{tab:cop.large}
\end{table}

    Table \ref{tab:cop.large} details the proportion of simulation repetitions in which the estimated posterior probability is largest for each data-generating and candidate copula combination. Both Figure \ref{fig:cop.large} and Table \ref{tab:cop.large} confirm that the posterior probabilities are well calibrated, and we can reliably distinguish between the various dependence structures in this large-sample setting. 

% \bibliographystyle{chicago}

% \bibliography{cobra}